\documentclass[12pt]{iopart}
\usepackage{epsfig}
\usepackage{iopams}
\bibstyle{unsrt}
\newcommand{\be}{\begin{equation}}
\newcommand{\ee}{\end{equation}}
\newcommand{\bea}{\begin{eqnarray}}
\newcommand{\eea}{\end{eqnarray}}

\newcommand{\dd}{\rmd}
\newcommand{\dyad}[1]{\buildrel{\leftrightarrow}\over{\mathbf{#1}}}
\newcommand{\mb}[1]{\bi{#1}}
\newcommand{\rot}{\mathrm{Curl}}
\newcommand{\im}[1]{\Im\mathrm{m}\left\{ #1 \right\}}

\newcommand{\vkl}{\mb{k}_\perp}

\newcommand{\kl}{k_\perp}
\newcommand{\sEM}{{\sum_{q=\mathrm{TE}}^\mathrm{TM}}}

\newcommand{\ki}{\kappa_1}
\newcommand{\kii}{\kappa_2}
\newcommand{\kg}{\kappa_g}
\newcommand{\Diq}{\Delta_{1q}}
\newcommand{\Diiq}{\Delta_{2q}}
\newcommand{\DiTE}{\Delta_{1}^\mathrm{TE}}
\newcommand{\DiTM}{\Delta_{1}^\mathrm{TM}}

\newcommand{\infint}{\int_0^\infty}
\newcommand{\matSum}{{\sum_{m=0}^\infty }'}
\newcommand{\rmL}{\mathrm{L}}
\newcommand{\FL}{\mathcal{F}_\mathrm{L}}

\begin{document}
\title[Casimir force on slab in cavity]{Casimir force on real materials - the slab and cavity geometry}
\author{Simen A Ellingsen\footnote{Current address: Department of War Studies, King's College London, Strand, London WC2R 2LS, UK} and
Iver Brevik }
\address{Department of Energy and Process
Engineering}
\address{Norwegian University of Science and
Technology, N-7491, Trondheim, Norway}
\ead{\emph{iver.h.brevik@ntnu.no}}

\date\today

\begin{abstract}
We analyse the potential of the geometry of a slab in a planar cavity for the purpose of Casimir force experiments.
 The force and its dependence on temperature, material properties and finite slab thickness are investigated both analytically and
  numerically for slab and walls made of aluminium and teflon FEP respectively. We conclude that such a setup is ideal
   for measurements of the temperature dependence of the Casimir force. By numerical calculation it is shown that temperature
    effects are dramatically larger for dielectrics, suggesting that a dielectric such as teflon FEP whose properties vary
     little within a moderate temperature range, should be considered for experimental purposes. We finally discuss the subtle
      but fundamental matter of the various Green's two-point function approaches present in the literature and show how they
       are different formulations describing the same phenomenon.
\end{abstract}

\maketitle

\section{Introduction}
\label{sec:intro}

The Casimir effect \cite{casimir48} can be seen as an effect of
 the zero point energy of vacuum which emerges due to the
 non-commutativity of quantum operators upon quantisation of the
 electromagnetic (EM) field. Although formally infinite in magnitude,
 the EM field density in bulk undergoes finite
 alterations when dielectric or metal boundaries are introduced in
 the system, giving rise to finite and measurable forces. As is
 well known, at nanometre to micrometre separations the Casimir
 attraction between bodies becomes significant, and the effect has
 attracted much attention during the last decade in the wake of
 the rapid advances in nanotechnology. The existence of the
 Casimir force was shown experimentally as early as 1958 by
 Spaarnay \cite{spaarnay58}, yet only recently new and much more
 precise measurements of Lamoreaux and others (see the review
 \cite{lamoreaux05}) have boosted the interest in the effect from a
 much broader audience. Experiments like that of Mohideen and Roy
 \cite{mohideen98}, and the very recent one of Harber {\it et.
 al.} \cite{harber05}, making use of the oscillations of a magnetically trapped Bose
 Einstein condensate, were subject to widespread regard. The same was true
 for the nonlinear micromechanical Casimir oscillator experiment
 of Chan {\it et al.} \cite{chan01,chan01a}.

 Recent reviews on the Casimir effect are given in
 Refs.~\cite{lamoreaux05,milton01,bordag01,milton04, nesterenko04}. Much
 information about recent developments can also be found in the
 special issues of {\it J. Phys. A: Math. Gen.} (May 2006)
 \cite{jphysa06}, and of {\it New J. Phys.} (October 2006)
 \cite{njp06}.

 Actual calculations of Casimir forces are usually performed via
 two different routes \cite{milton01}; either by summation of the
 energy of discrete quantum modes of the EM field
 (cf., for instance,  \cite{plunien86}), or via a Green's function method first
 developed by Lifshitz \cite{lifshitz56}. Mode summation, despite
 its advantage of a simpler and more transparent formalism, is
 usually far inferior. In practice it is only in systems where
 quantum energy states are known that energy summation can be
 carried out explicitly. This requires the system to be highly
 symmetric, and favour assumptions such as perfectly conducting walls like in
 the original Casimir problem. Geometries in which quantum states
 are known exactly, unfortunately, are few.

 The method of calculating the force through Green's functions
 avoids some but not all of these problems; exact solutions are
 still only known in highly symmetrical systems such as infinitely
 large parallel plates or concentric spheres. Via the
 fluctuation-dissipation theorem, the EM field energy
 density is linked directly to the photonic Green's function, and
 the force surface density acting on boundaries can be
 calculated, at least in principle. The theory of Green's
 functions and the application of them will be central in the
 present paper.

 The purpose of the present work is twofold. First, we intend to
explore some of the delicate issues that occur in the
 Green's function formalism in typical settings involving
 dielectric boundaries. Upon relating the two-point
 functions to the Green's function one may choose to calculate the
 Green function in full \cite{milton01,hoye03}. The method is
 complete but may appear cumbersome, at least so in the presence
 of several dielectric surfaces. It is possible to reduce the
 calculational burden somewhat by simplifying the Green function
 expressions,   by omitting those parts that do not contribute to
 the Casimir force. This means that one works with ``effective'' Green
 functions. This method is employed and briefly discussed by Lifshitz and co-workers, cf.\ e.g.\
 \cite{lifshitz80}. The connections between the different
 kinds of Green's functions are in our opinion far from trivial,
 and we therefore believe it of interest to present some of
 the formulas that we have compiled and which have turned out to be useful in
 practice.

 As for the calculational technique for the Casimir force in a
 multilayer system, there exists a powerful formalism worked out,
 in particular, by Toma\v{s} \cite{tomas95}. In
 turn, this formalism was based on work by Mills and Maradudin two
 decades earlier \cite{mills75}. One of us recently made a review
 of this technique, with various applications \cite{ellingsen06}.
 We shall make use of this technique in the following. In company
 with the by now classic theory of Lifshitz and co-workers
 \cite{lifshitz56,dzyaloshinskii61} and the standard Fresnel theory in optics,
 the necessary set of tools is provided.

Our second purpose is to apply the formalism to concrete
calculations of the Casimir pressure on a dielectric plate in a
multilayer setting. Especially, we will consider the pressure on a
plate situated in a cavity (5-zone system). We work out force
expressions and eigenfrequency changes when the plate is acted
upon by a harmonic-oscillator mechanical force (spring constant
$k$) in addition to the Casimir force, and is brought to oscillate
horizontally. To our knowledge, explicit calculations of this sort
have not been made before. A
chief motivation for this kind of calculation is that we wish to
evaluate the magnitudes of {\it thermal corrections} to the
Casimir pressure. In recent years there have been lively
discussions in the literature about the thermal corrections; for
some statements of both sides of the controversy, see
Refs.~\cite{hoye03,decca05,brevik05,bentsen05,bezerra06,hoye06,mostepanenko06,brevik06,brevik06a}.
We hope that the consideration of planar multilayer systems may
provide additional insight into the temperature problem.

We will be considering  {\it uniformly heated systems} only. The
recent experiment of Harber {\it et al.} \cite{harber05}
investigated the surface-atom force at thermal equilibrium at room
temperature, the goal being to measure the  surface-atom force at
very large distances, taking into account the peculiar properties
of a Bose-Einstein condensate gas. Later, the same group
investigated the non-equilibrium effect \cite{obrecht06}. This
paper seems to have reported the first accurate measurement of the
thermal effect (of any kind) of the Casimir force, in good
agreement with earlier theoretical predictions \cite{antezza05}
 (cf. also the prior theory of Pitaevskii on the non-equilibrium
dynamics of EM fluctuations \cite{pitaevskii00}). Consideration of
such systems lies, however, outside the scope of the present
paper.

The following point ought also to be commented upon, although it
is not a chief ingredient of the present paper: Our problem bears
a relationship to the famous Abraham-Minkowski
controversy, or more generally the question of how one should construct the
correct form of the EM energy-momentum tensor in a
medium. This problem has been discussed more and less
intensely ever since Abraham and Minkowski proposed their
energy-momentum expressions around 1910. The advent of accurate
experiments, in particular, has aided a better insight into
this complicated aspect of field-matter interacting systems. Some years ago,
one of the present authors  wrote a review of the experimental
status in the field \cite{brevik79} (cf. also \cite{brevik86}).
There is by now a rather extensive literature in this field; some papers
are listed in
\cite{moller72,kentwell87,antoci98,obukhov03,loudon97,
garrison04,feigel04,leonhardt06}. In the present case, where the
EM surface force on a dielectric boundary results
from integration of the volume force density across the boundary
region, the Abraham and Minkowski predictions actually become
equal. Recently, in a series  of papers Raabe and Welsch have expressed
the opinion that the Abraham-Minkowski theory is inadequate and
that a different form of the EM energy-momentum
tensor has to be employed
\cite{raabe05,welsch05,raabe05a,raabe06}. We cannot agree with this
statement, however. All the experiments in optics that we are
aware of can be explained in terms of the Abraham-Minkowski theory
in a straightforward way. One typical example is provided, for
instance, by the oscillations of a water droplet illuminated by a
laser pulse. Some years ago, Zhang and Chang made an experiment in
which the oscillations of the droplet surface were clearly
detectable \cite{zhang88}. It was later shown theoretically how
the use of the Abraham-Minkowski theory could reproduce the
observed results to a reasonable accuracy \cite{lai89,brevik99}.
In our theory below, we will use the Abraham-Minkowski theory
throughout.

SI units are used throughout the calculations, and permittivity
$\epsilon$ and permeability $\mu$ are defined as relative
(nondimensional) quantities. We thus write ${\bf D}=\epsilon_0
\epsilon {\bf E}$, ${\bf B}=\mu_0 \mu {\bf H}$.

The outline of the paper is as follows. In the next section we
analyse the 5-layered magnetodielectric system (figure 1),
presenting the full Green's function as well as its effective (or
reduced) counterpart. We here aim at elucidating some points in
the formalism that in our opinion are rather delicate. Section 3
is devoted to a study of an oscillating slab in a Casimir cavity,
permitting,  in principle at least, how the change in the
eigenfrequency of the slab with respect to the temperature can
give us information about the temperature dependence of the
Casimir force. Section 4 discusses more extensively the
relationships between the Green's two-point functions as
introduced by Lifshitz  et al., and by Schwinger  et al. In
section 5 we present results of numerical Casimir force
calculations for selected substances, taking Al as example of a
metal, and teflon FEP as example of a dielectric\footnote{As a word of
caution, we mention here that our permittivity data for metals are
intended to hold in the bulk, whereas in practice the real and
imaginary parts of the index of refraction for metals come from
ellipsometry measurements, and are thus really {\it surface}
measurements. There is an inherent uncertainty in the calculated
results coming from this circumstance, of unknown magnitude,
although in our opinion the corrections will hardly exceed the 1\%
level due to the general robustness of the force expression
against permittivity variations. Ideally, information about the
permittivity versus imaginary frequency would be desirable, for a
metallic film. We thank Steve Lamoreaux for comments on this
point.}. In section 6 we consider the effect of finite slab
thickness, i.e. the ``leakage'' of vacuum radiation from one gap
to the other. We find the striking result that for dielectrics the
relative finite thickness correction is much larger than for
metals. For teflon FEP versus Al the relative correction is almost
two orders in magnitude higher.

A word is called for, as regards the permeability $\mu$. As
anticipated above, we allow $\mu$ to be different from 1. This is
motivated chiefly by completeness, and is physically an
idealization. It is known that the permeability for most materials
is lossy at high frequencies, corresponding to imaginary values
for $\mu$. That phenomenon is limited to a restricted frequency
interval, however,  (10 - 100 GHz), and loses effect at the higher
frequencies.

%%%%%%%%%%%%%%%%%%%%%%% SECTION %%%%%%%%%%%%%%%%%%%%
\section{Casimir force on a slab in a cavity}
\label{Sec:5-zone}

\begin{figure}
  \begin{center}
  \includegraphics[width=3.5in]{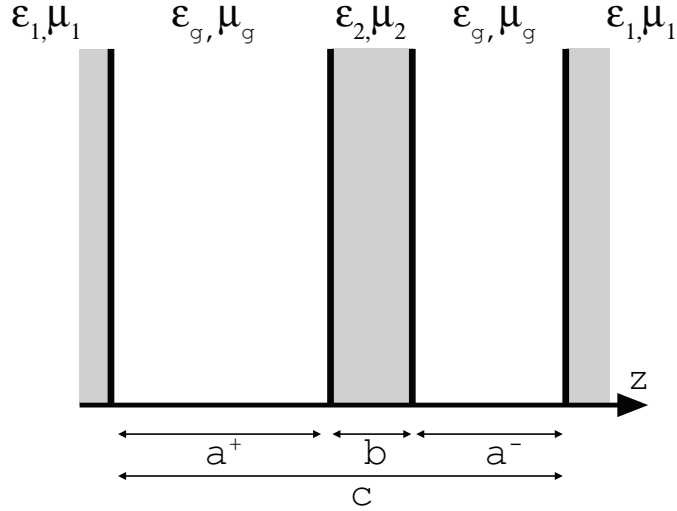}
  \caption{The five-zone geometry of a slab in a cavity. We have chosen $z=0$ at the left hand cavity wall.}\label{fig_5z}
  \end{center}
\end{figure}

We shall consider a 5-layered magnetodielectric system such as depicted in figure \ref{fig_5z}. The analytical calculation of the Casimir force density acting on the slab in such a geometry is well known; it may be calculated, quite simply, by a straightforward generalisation of the famous calculation by Lifshitz and co-workers used for the simpler, three-layered system of two half-spaces separated by a gap \cite{lifshitz80,dzyaloshinskii61}.

Rather than starting from the photonic Green's function as a
propagator as known from quantum electrodynamics, we introduce
classical and macroscopic two-point (Green's) function according
to the convention of Schwinger et.al.\ \cite{schwinger78} as
\be\label{def_Gamma}
  \mb{E}(x) = \frac{1}{\epsilon_0}\int \dd^4x' \dyad{\Gamma}(x,x')\cdot \mb{P}(x'),
\ee
where $x=(\mb{r},t)$. Due to causality, $t'$ is only integrated over the region $t'\leq t$. It follows from Maxwell's equations that $\mathbf{\Gamma}$ obeys the relation
\be\label{eq_basic_Gamma}
  \nabla\times\nabla\times \dyad{\Gamma}(\mb{r},\mb{r}'; \omega) - \frac{\epsilon(\mb{r})\mu(\mb{r})\omega^2}{c^2}\dyad{\Gamma}(\mb{r},\mb{r}'; \omega) = \frac{\mu(\mb{r})\omega^2}{c^2}\delta(\mb{r}-\mb{r}')\dyad{1},
\ee
where we have performed a Fourier transformation according to
\be \label{eq_Fourier_w}
  \dyad{\Gamma}(x,x') = \int_{-\infty}^\infty \frac{\dd \omega}{2\pi}\rme^{-i\omega \tau}\dyad{\Gamma}(\mb{r}, \mb{r}';\omega),
\ee
with $\tau \equiv t-t'$.

A comparison of (\ref{eq_basic_Gamma}) with the corresponding equation in \cite{lifshitz80,dzyaloshinskii61} shows formally that
 $\Gamma$ is essentially equivalent with the retarded photonic Green's function in a medium\footnote{Compared to Lifshitz et.al.\ $\mathcal{D}=-\hbar c^2 \Gamma / \omega^2$, which is only a matter of definition.}. The physical connection is not entirely trivial, however. As motivation we notice that (\ref{def_Gamma}) expresses the linear relation between the dipole density at $x'$ and the resulting electric field at $x$, in essence the extent to which an EM field is able to \emph{propagate} from $x'$ to $x$. This is exactly the classical analogy of the quantum definition of a Green's function propagator, in accordance with the correspondence principle as introduced by Niels Bohr in 1923. We note furthermore that insisting that $t'\leq t$ ensures that account is taken of retardation, corresponding to the Lifshitz definition of the retarded photonic Green's function (e.g.\ \cite{lifshitz80} \S75) which is 0 for $t<t'$.

We make use of the Fluctuation-Dissipation theorem at zero temperature, rendered conveniently as
\numparts
\begin{eqnarray}
  \rmi\langle E_i(\mb{r})E_k(\mb{r}')\rangle_\omega &= \frac{\hbar}{\epsilon_0}\im{\Gamma_{ik}(\mb{r}, \mb{r}'; \omega)}\label{eq_EE}\\
  \rmi\langle H_i(\mb{r})H_k(\mb{r}')\rangle_\omega &= \frac{\hbar}{\mu_0}\frac{c^2}{\mu\mu'\omega^2}\rot_{ij}\rot_{kl}'\im{\Gamma_{jl}(\mb{r}, \mb{r}'; \omega)}\label{eq_HH},
\end{eqnarray}
\endnumparts
with the notation $\rot_{ik}\equiv \epsilon_{ijk}\partial_j$ ($\epsilon_{ijk}$ being the Levi-Civita symbol and summation over identical indices is
 implied), $\rot_{ik}' \equiv \epsilon_{ijk}\partial_j'$ where $\partial_j'$ is differentiation with respect to component $j$ of $\mb{r}'$,
  and $\mu'\equiv \mu(\mb{r}')$. The brackets denote the mean value with respect to fluctuations. The Casimir pressure acting on some surface is now
   given by the $zz$-component of the Abraham-Minkowski stress tensor, found by simple insertion to become\cite{lifshitz80,dzyaloshinskii61}\footnote{The expression is generalised compared to the original reference to allow $\mu\neq 1$.}
\be \label{eq_Tzz_Lifshitz}
  \mathcal{F}_z = \hbar \int_0^\infty \frac{\dd \zeta}{2\pi}\left[\epsilon(\Gamma_{xx}^{E}+\Gamma_{yy}^{E}-\Gamma_{zz}^{E})+\frac{1}{\mu}(\Gamma_{xx}^{H}+\Gamma_{yy}^{H}-\Gamma_{zz}^{H}) \right]_{\mb{r}=\mb{r}'},
\ee
where a standard frequency rotation $\omega = \rmi\zeta$ has been performed and the convenient quantities $\Gamma^E$ and $\Gamma^H$ have been defined according to
\numparts
\begin{eqnarray}
  \Gamma^E_{ik}(\mb{r}, \mb{r}'; \omega) &\equiv& \Gamma_{ik}(\mb{r}, \mb{r}'; \omega);\\
  \Gamma_{ik}^H(\mb{r}, \mb{r}'; \omega) &\equiv& \frac{c^2}{\omega^2}\rot_{il}\rot_{km}'\Gamma_{lm}(\mb{r}, \mb{r}'; \omega).
\end{eqnarray}
\endnumparts

In (\ref{eq_Tzz_Lifshitz}) only the homogeneous (geometry dependent) solution of (\ref{eq_basic_Gamma}) is included; the inhomogeneous solution pertaining to the delta function represents the solution inside a homogeneous medium filling all of space. This term is geometry independent, and cannot contribute to any physically observable quantity. Importantly, however, any such simplification from the full Green's function to its ``effective'' counterpart must only be made subsequent to all other calculations.

The system is symmetrical with respect to translation and rotation in the $xy$-plane and we transform the Green's function once more:
\be
  \dyad{\Gamma}(\mb{r},\mb{r}';\omega)= \int \frac{\dd^2 \kl}{(2\pi)^2}\rme^{i\vkl\cdot(\mb{r}_\perp-  \mb{r}_\perp')}\dyad{g}(z,z';\vkl,\omega).
\ee
Here and henceforth, the subscript $\perp$ refers to a direction in the $xy$-plane. In the $\vkl,\omega$ Fourier domain one finds \cite{milton01,schwinger78} that the component equations (\ref{eq_basic_Gamma}) combine to (among others) the equations
\be\label{gxx}
  (\partial_z^2 - \kappa^2)g_{xx}(z,z';\vkl,\omega) = \frac{\kappa^2}{\epsilon\mu}\delta(z-z')
\ee
and
\be\label{gyy}
  (\partial_z^2 - \kappa^2)g_{yy}(z,z';\vkl,\omega) = -\frac{\mu\omega^2}{c^2}\delta(z-z'),
\ee
which readily give us these two components in each homogeneous zone. We have defined the quantity $\kappa \equiv (k_\perp^2 - \epsilon\mu \omega^2/c^2)^{1/2}$. The final diagonal component is found by means of the relations
\numparts
\begin{eqnarray}
  g_{zz}(z,z';\vkl,\omega) &=& -\frac{\rmi \kl}{\kappa^2}\partial_z g_{xz}(z,z';\vkl,\omega) + \frac{1}{\kappa^2}\frac{\mu\omega^2}{c^2}\delta(z-z') \\
  g_{zx}(z,z';\vkl,\omega) &=& -\frac{\rmi \kl}{\kappa^2}\partial_z g_{xx}(z,z';\vkl,\omega) \\
  g_{xz}(z,z';\vkl,\omega) &=& g_{zx}(z',z; -\vkl, \omega),
\end{eqnarray}
\endnumparts
of which the first two are components of (\ref{eq_basic_Gamma}) and the last was shown by Lifshitz (e.g. \cite{lifshitz80}).

We return to the geometry of figure \ref{fig_5z}. An important point to emphasize is that unlike certain authors in the past (e.g.\ \cite{matloob01}) we make no principal difference between the walls of the cavity and the slab; they are both made of real materials with finite permittivity and conductivity at all frequencies as is the case in any real experimental setting. The net force density per unit transverse area acting on the slab is found by first placing the source (i.e.\ $z'$) in one of the gaps and calculate the resulting Green's function in this gap. This yields the attraction the stack of layers to the left and right of this gap exert upon each other. The procedure is then repeated with respect to the other gap region and the net force on the slab found as the difference between the two.

The solution of (\ref{gxx}) may be written down directly, yielding in the case where $z'$ lies in the gap region to the left of the slab
\begin{equation}
  g_{xx} = \left\{
    \begin{array}{lc}
      A\rme^{\ki z} & z<0 \\
      C_1\rme^{\kg z}+C_2\rme^{-\kg z} + G \rme^{-\kg |z-z'|} &0<z<a^+\\
      E_1\rme^{\kii z}+ E_2\rme^{-\kii z} &a^+<z<a^++b\\
      D_1\rme^{\kg z}+D_2\rme^{-\kg z} &a^++b<z<c \\
      B \rme^{-\ki z} & z>c
    \end{array} \right.
\end{equation}
with $G=-\kg/(2\epsilon_g \mu_g)$. The ``constants'' $A$ to $E$ are $z'$-dependent. From standard conditions of EM field continuity and (\ref{eq_EE},{\it b}) one may show that $g_{xx}$ and $(\epsilon/ \kappa^2) \partial_z g_{xx} $ are continuous across interfaces, giving a total of 8 equations which are solved with respect to $C_1$ and $C_2$ yielding after lengthy but straightforward calculation (for details, cf.\ \cite{ellingsenMaster}) the solution in the left hand gap ($0<z<a^+$, denoted with superscript $+$)\label{eq_greens_complete}
\[
    \fl g_{xx}(+) = -\frac{\kappa_g}{2\epsilon_g} \left\{ \frac{1}{d^+_\mathrm{TM}}\left[2\cosh\kg(z-z') + \frac{e^{\kg(z+z')}}{\DiTM}+\DiTM e^{-\kg(z+z')} \right] + \DiTM e^{-\kg(z+z')}\right\}.
\]
Here and henceforth the inhomogeneous $|z-z'|$-term has been omitted subsequent to other calculation as argued above. Foreknowingly,  we have defined the key quantities
\be \label{d}
  \frac{1}{d^\pm_q} = \frac{U_q^\mp \rme^{-2\kg a^\pm}}{V_q^\mp-U_q^\mp \rme^{-2\kg a^\pm}};\hspace{20pt} q=\{\mathrm{TE,TM}\}
\ee
where
\begin{eqnarray*}
  U_q^\pm &=& \Diq\Diiq (1-\Diq\Diiq \rme^{-2\kg a^\pm}) - \Diq(\Diiq - \Diq \rme^{-2\kg a^\pm})\rme^{-2\kii b},\\
  V_q^\pm &=& 1-\Diq\Diiq \rme^{-2\kg a^\pm} - \Diiq(\Diiq - \Diq \rme^{-2\kg a^\pm})\rme^{-2\kii b},
\end{eqnarray*}
using the single-interface Fresnel reflection coefficients
\be \label{eq_Delta_def}
  \Delta_{i,q} = \frac{\kappa_i-\gamma_{i,q}\kappa_g}{\kappa_i+\gamma_{i,q}\kappa_g};\hspace{20pt} \gamma_{i,q} = \left\{\begin{array}{cl}\mu_i/\mu_g, & q=\mathrm{TE}\\ \epsilon_i/\epsilon_g, &q=\mathrm{TM}\end{array}\right., \hspace{20pt} i = 1,2.
\ee
Note already how the quantity $(d^\pm)^{-1}$ is a generalisation of the quantity $d^{-1}$ as it was defined for the three-layer system by Schwinger et.al.\ \cite{milton01,schwinger78} (dubbed $\Delta$ in the Lifshitz et.al.\ literature). In the limit $\kappa_2\to\kg$ we immediately get $(d_q^\pm)^{-1} \to (\Diq^{-2}e^{2\kg c}-1)^{-1}$, i.e.\ the three-layer standard result for a cavity of width $c$ with no slab.

Following the above described procedure we get
\[
  \fl g_{zz}(+) = \frac{k_\perp^2}{2\kappa_g\epsilon_g} \left\{ \frac{1}{d^+_\mathrm{TM}}\left[2\cosh\kg(z-z') - \frac{e^{\kg(z+z')}}{\DiTM}-\DiTM e^{-\kg(z+z')} \right] + \DiTM e^{-\kg(z+z')}\right\} .
\]

Exactly the same procedure as for $g_{xx}$ is followed to obtain the $yy$-component. One finds that $g_{yy}$ and $\mu^{-1}\partial_zg_{yy}$ are continuous across boundaries, giving 8 new equations solved as above to yield
\[
  \fl g_{yy}(+) = \frac{\mu_g}{2\kg}\frac{\omega^2}{c^2}\left\{ \frac{1}{d^+_\mathrm{TE}}\left[2\cosh\kg(z-z') - \frac{e^{\kg(z+z')}}{\DiTE}-\DiTE e^{-\kg(z+z')} \right] - \DiTE e^{-\kg(z+z')}\right\} .
\]

The results for the right hand ($-$) gap is found by transforming the above results according to $a^\pm\to a^\mp$ and $z\to c-z$.

To obtain the force density on each side of the slab, the solutions are now inserted into (\ref{eq_Tzz_Lifshitz}). One may show \cite{ellingsenMaster} that the terms depending on $z+z'$ do not contribute to the force density (this is a subtle point which will be discussed further below). Upon omitting these terms, the right hand expressions are simply given by swapping $+$ and $-$ indices everywhere and we are left with the \emph{effective} Green's function solution in the $\omega,\kl$-domain:
\numparts
\begin{eqnarray}
  g_{xx}(\pm) = -\frac{\kappa_g}{\epsilon_g} \frac{1}{d^\pm_\mathrm{TM}}\cosh\kg(z-z') \\
  g_{yy}(\pm) = \frac{\omega^2\mu_g}{c^2\kg}\frac{1}{d^\pm_\mathrm{TE}}\cosh\kg(z-z') \\
  g_{xx}(\pm) = \frac{k_\perp^2}{\kg\epsilon_g} \frac{1}{d^\pm_\mathrm{TM}}\cosh\kg(z-z') .
\end{eqnarray}
\endnumparts

Upon insertion into (\ref{eq_Tzz_Lifshitz}) we find the force on either side of the slab yielding the net Casimir pressure acting on the slab towards the right as
\be\label{eq_5z_F(a)}
  \mathcal{F}^0(a^+,a^-;b,c) = \frac{\hbar}{2\pi^2}\infint \rmd \zeta \infint \rmd \kl\cdot \kl\kg \sEM \left(\frac{1}{d_q^-}-\frac{1}{d_q^+}\right).
\ee
Naturally, the force will always point away from the centre position. Superscript $0$ here denotes that the expression is taken at zero temperature. The finite temperature expression, as is well known, is found by replacing the frequency integral by a sum over Matsubara frequencies according to the transition
\[
  \hbar\infint \frac{\rmd \zeta}{2\pi} f(\rmi\zeta) \to k_B T\matSum f(\rmi\zeta_m);\hspace{20pt} \rmi\zeta_m = \rmi(2\pi k_BT/\hbar)\cdot m
\]
yielding
\be \label{eq_5z_F(aT)}
  \mathcal{F}^T(a^+,a^-;b,c) = \frac{k_BT}{\pi}\matSum \infint \rmd \kl\cdot \kl\kg \sEM \left(\frac{1}{d_q^-}-\frac{1}{d_q^+}\right).
\ee
The prime on the summation mark denotes that the zeroth term is given half weight as is conventional.

Rather than painstakingly solving the eight continuity equations to obtain the Green's function as above, the result (\ref{eq_5z_F(a)}) is found much more readily using a powerful procedure following Toma\v{s} as presented recently by one of us \cite{ellingsen06}. The above result was obtained by Toma\v{s} \cite{tomas02} presumably using this procedure. It was worth going through the above calculations, however, for the sake of shedding light on some in our opinion non-trivial details which are often tacitly bypassed.

%%%%%%%%%%%%%%%%%%%%%%%%%%%%%%% SECTION %%%%%%%%%%%%%%%%%%%%%%%%%%%%%%
\section{Casimir measurement by means of an oscillating slab}
\label{ch_oscillate}

Equation (\ref{eq_5z_F(aT)}) may be written on a more handy form in terms of the distance $\delta$ from the centre of the slab to the midline of the cavity. We introduce the system parameter $h=c-b=a^++a^-$ and substitute according to $a^\pm = h/2 \pm \delta$. With some straightforward manipulation we are able to write (\ref{eq_5z_F(aT)}) as
\be\label{eq_5z_F(delta)}
  \mathcal{F}^T(\delta; b,c) = \frac{k_BT}{\pi}\matSum \int_0^\infty \dd\kl\cdot \kl\kg \sEM \frac{A_q\sinh 2\kg \delta}{B_q-A_q \cosh 2\kg\delta}
\ee
with
\begin{eqnarray*}
  A_q &=& 2\Diq\Diiq(1-\rme^{-2\kii b})\rme^{-\kg h},\\
  B_q &=& 1-\Diiq^2\rme^{-2\kii b} + \Diq^2 (\Diiq^2 - \rme^{-2\kii b})\rme^{-2\kg h}.
\end{eqnarray*}

\begin{figure}
  \begin{center}
    \includegraphics[width=2.5in]{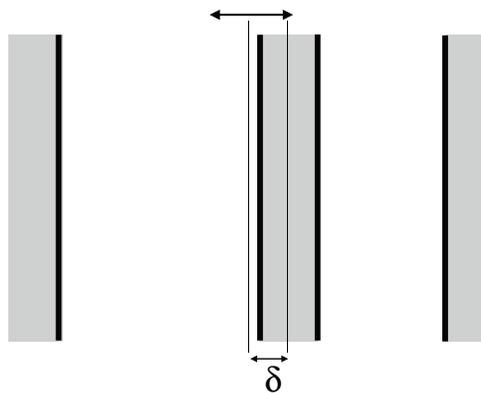}
    \caption{The slab oscillates about the cavity midline. We imagine a spring is attached to the slab exercising a Hooke-force towards the equilibrium position.}
  \end{center}
\end{figure}

We write the force on the slab at finite temperatures as a Taylor expansion to first order in $\delta$ as
\be\label{eq_Taylor}
  \mathcal{F}^T(\delta; b,c) = a_1 \delta + \mathcal{O}(\delta^3)
\ee
with
\be
  a_1(T; b,c) = \frac{2k_BT}{\pi}\matSum \infint \rmd k_\perp\cdot \kl \kappa_g^2 \sEM \frac{A_q}{B_q-A_q}.
\ee

Assume now the slab is attached to a spring with spring constant $k$ per unit transverse area. For small $\delta$ we may assume the slab to oscillate in a harmonic fashion (assuming $k>a_1$ now) with frequency given by Newton's second law as
\[
  \Omega = \Omega_0 - \Delta\Omega(T) = \sqrt{\frac{k-a_1(T)}{m}},
\]
where $\Omega_0 = \sqrt{k/m}$ and $m$ is the mass of the slab per unit transverse area. In the case that $k\gg a_1$ we get
\[
  \Delta\Omega(T) \approx \frac{a_1(T)}{2\sqrt{km}} = \Omega_0 \frac{a_1(T)}{2k}.
\]
We show by numerical calculation in chapter \ref{ch_numerics} how the Taylor coefficient $a_1(T)$ varies significantly with $T$ rendering an oscillating slab-in-cavity setup possibly suitable for future experimental investigation of the true temperature dependence of the Casimir force.

The setup as described is somewhat reminiscent of the setup currently employed by Onofrio and co-workers in Grenoble \cite{lambrecht05} where plates mounted on a double torsion balance are attracted to a pair of fixed plates. In their planned experiment, the distance from plate to wall will however kept constant during force measurements. Indeed, a double torsion balance might be one way of envisioning an experimental realisation essentially equivalent to the system described (if thickness corrections are neglected) if the plates are mounted such that when one pair of plates approach each other, separation is increased between the pair on the opposite side of the pendulum. An even closer relative might be the recent experiments in Colorado where perturbations of the eigenfrequency of a magnetically trapped Bose-Einstein substrate in the vicinity of a surface provides a sensitive force measurement technique \cite{harber05,obrecht06,mcguirk04}. Both of these experiments involve a plate (in the widest sense) attracted to a wall on only \emph{one} side; an ``open cavity''. 

While a one-sided configuration is possibly experimentally simpler, there are two physical advantages of the sandwich geometry as presented here: the frequency shift $\Delta\Omega(T)$ is essentially twice as large using a closed cavity and, perhaps more importantly, in a symmetrical geometry the harmonical approximation ($\mathcal{F}^T\propto \delta$) is accurate for larger deviations $\delta$ from the equilibrium position than is the case for an open geometry. These points are elaborated further in \ref{appendix}.

%************* SECTION ******************
\section{Fundamental discussion: two-point functions and Green's functions}

In the standard Casimir literature there are two famous and somewhat different derivations of the classical Lifshitz expression\footnote{By ``Lifshitz force'' is henceforth meant the Casimir force between two plane parallel (magneto)dielectric half-spaces separated by a medium different from both. By the ``Lifshitz expression'' is meant the mathematical expression for this force as derived by Lifshitz and co-workers \cite{dzyaloshinskii61}.}, namely that of Lifshitz and co-workers in 1956-61 \cite{lifshitz56, lifshitz80} and that of Schwinger and co-workers some years later \cite{milton01, schwinger78}. The two both make use of a Green's two-point function but in two different ways which upon comparison seem somewhat contradictory at first glance. Understanding how they relate to each other is not trivial in our opinion.

In order to calculate the force acting on an interface between two different media, both schools calculate what in our coordinates is the $zz$ component of the Abraham-Minkowski energy momentum tensor as described above using the Green's function through the fluctuation-dissipation theorem as in (\ref{eq_EE},$b$). Lifshitz argues as recited above that in his formalism some terms of the Green's function (those dependent on $z+z'$) make no contribution to the force\footnote{This is shown formally in \cite{ellingsenMaster}.}. These are consequently omitted, leaving an \emph{effective} Green's function. Schwinger et.al., however, make use of the \emph{entire} Green's function ultimately arriving at an expression similar to (\ref{eq_Tzz_Lifshitz}) in which the $z+z'$ terms \emph{are} included and indeed necessary in order to reproduce Lifshitz' result. The $|z-z'|$ dependent source term is geometry independent and eventually omitted in both references.

To solve the paradox we recognise one important difference between the two procedures: Lifshitz takes the limit $\mb{r}\to\mb{r}'$ so that $\mb{r}$ and $\mb{r}'$ are both on the same side of one of the sharp interfaces, whereas in Schwinger's method, $\mb{r}$ is on one side whilst $\mb{r}'$ is on the other. By using continuity conditions for the EM field, calculations can be carried out with analytic knowledge of the Green's function only on one side of the interface in both cases, thus masking this principal difference. Remembering that $T_{zz}$ is the density of momentum flux in the $z$-direction, the physical difference between the methods is that whilst Lifshitz calculates the force density as the \emph{net} stream of momentum into one side of the interface, Schwinger et.al.'s expression represents the \emph{entire} stream into one side minus the entire stream out of the other side. Due to conservation of momentum, the procedures are physically equivalent.

The question remains how to interpret the terms dependent on $z+z'$. Arguably, the absolute value of such terms must be arbitrary, since they will depend on the position of an arbitrarily placed origin\footnote{The notion of arbitrarily large energy densities, of course, is not foreign to Casimir calculations; Casimir's original calculation involved the difference between the apparently infinite energy density of the zero-point photon field in the absence and presence of perfectly conducting interfaces.}. Furthermore, since these terms cancel each other perfectly in (\ref{eq_Tzz_Lifshitz}), one may think of them as representing an isotropic flux of photonic momentum, flowing in equal amounts in both directions along the $z$-axis, giving rise to no measurable effect \emph{inside a homogeneous medium}.

Schwinger, however, insists $\mb{r}$ and $\mb{r}'$ lie
infinitesimally close to \emph{either side} of an interface. While
the $z+z'$ terms cancel each other when all calculated in the same
medium, their values depend on $\epsilon$ and $\mu$, so when
$\epsilon\neq \epsilon'$ or $\mu\neq\mu'$, their net contribution
is finite.

This is exactly made up for in Lifshitz' approach by the fact that a sudden change in permittivity and permeability
(such as at an interface between a dilute and an opaque medium) causes some of the radiation to be reflected off the
interface in accordance with Fresnel's theory. Thus although $z$ and $z'$ both lie inside the same medium, there is a net
 flow of momentum either out of (attractive) or into (repulsive) the gap giving rise to a Casimir force.
Such an analysis of the use of Green's functions gives way for an
understanding of how three different representation of the Casimir
effect come together; the derivation by Lifshitz starting from
photonic propagators in quantum electrodynamics, that by Schwinger
et.al.\ based on Green's function calculations from classical
electrodynamics and a third approach based on Fresnel theory which
we may refer to as the ``optical approach'' (originally in form of
non-retarded Van der Waals theory \cite{langbein74,parsegian06},
recently revisited by Scardicchio and Jaffe, see
\cite{scardicchio04} and references therein).

\begin{figure}
  \begin{center}
    \includegraphics[width=4in]{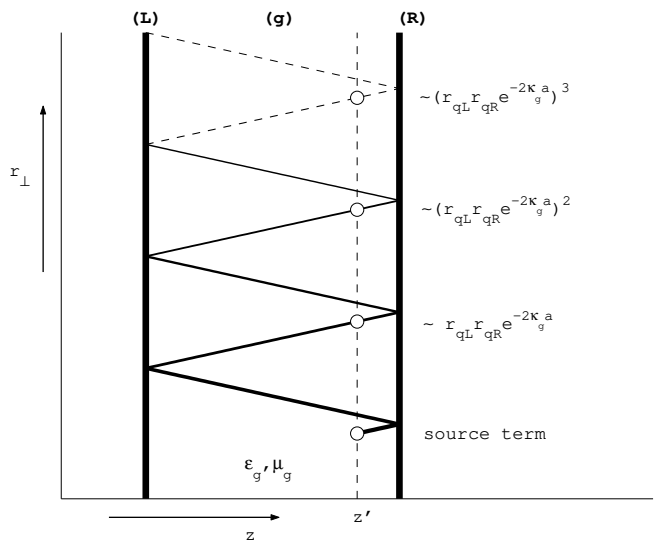}
    \caption{Contributions to $\dyad{g}$ in a gap between two bodies in the optical visualisation. The distance between the bodies is $a$. Each term has a weight factor as shown on the right hand side. The sum of the infinitely many reflections of a $q$-polarised wave is $d_q^{-1}$.}
    \label{fig_refl}
  \end{center}
\end{figure}

We showed that the factors $(d_q^\pm)^{-1}$ were generalised versions of the factors denoted $d^{-1}$ and $(d')^{-1}$ in Schwinger et.al.'s theory for the three-layer model. These are both special cases of a more general quantity
\[
  \frac{1}{d_q}=\frac{r_{q\mathrm{L}}r_{q\mathrm{R}}\rme^{-2\kg a}}{1-r_{q\mathrm{L}}r_{q\mathrm{R}}\rme^{-2\kg a}}
\]
pertaining to a gap of width $a$ separating planar bodies to the
left (L) and right (R) of it whose Fresnel reflection coefficients
are $r_{q\mathrm{L}}$ and $r_{q\mathrm{R}}$ respectively. If the
media are infinitely large and homogeneous media indexed 1 and 2
respectively, say, $r_{q\mathrm{L}}$ and $r_{q\mathrm{R}}$ are
simply $-\Diq$ and $-\Diiq$ from (\ref{eq_Delta_def}); if the
bodies are more complex, e.g.\ has a multilayered structure, their
corresponding Fresnel coefficients
 will be more complicated. This is discussed in detail in \cite{ellingsen06}. An EM plane wave with momentum
 $\hbar \mathbf{k}$ is described as $\rme^{\rmi(\mathbf{k}_\perp\cdot\mathbf{r}_\perp + k_zz)}$. In medium $g$,
 furthermore, $k_z=\rmi\kg$ according to Maxwell's equations, i.e.\ the wave is evanescent in the $z$-direction if
 $k_\perp^2>\epsilon_g\mu_g\omega^2/c^2$ (otherwise propagating). After frequency rotation $\omega^2\to-\zeta^2$ this is always
  true ($k_\perp$ is assumed real), so \emph{every} wave is described as an evanescent wave.  The attenuation
of an EM field of frequency $\rmi\zeta$ propagating a
 distance $l$ along the $z$-axis in medium $g$ is $\exp(-\kg l)$, so one readily shows that $d_q^{-1}$ is the sum of \emph{relative amplitudes} of the electric fields having travelled all paths starting and ending at the same $z$-coordinate and with the same direction:
\[
  \frac{1}{d_q} = r_{q\mathrm{L}}r_{q\mathrm{R}}\rme^{-2\kg a} + (r_{q\mathrm{L}}r_{q\mathrm{R}}\rme^{-2\kg a})^2 + ...=\sum_{n=1}^\infty (r_{q\mathrm{L}}r_{q\mathrm{R}}\rme^{-2\kg a})^n.
\]
An illustration of this is found in figure \ref{fig_refl}. Since the phase shift from propagation in the $\perp$ direction is disregarded in this respect, one might think of $d_q^{-1}$ as a sum over all \emph{closed} paths, parallel to the $z$-axis and starting and ending in the same point.

Considering again the expressions for the complete Green's
functions $g_{xx}, g_{yy}$ and $g_{zz}$ in section 2,
 we see that the last term of all three components are the only ones not multiplied by a factor $d_q^{-1}$ (indices $\pm$ suppressed). Since this factor is the only part of $\dyad{g}$ containing geometry information, the last term is geometry independent, and can obviously make no contribution to a physical force. Hence: all contributing terms are proportional with $d_q^{-1}$ which leads us to the conclusion that the Casimir attraction between bodies on either side of a gap region at a given temperature depends solely on the extent to which some EM field originating in the gap, stays in the gap.

To sum it all up, we argued that Schwinger's classical Green's function as introduced is the exact macroscopic analogy of
 Lifshitz' QED propagator according to Bohr's correspondence principle. In its Fourier transformed form it expresses the
  probability amplitude that an electric field which has transverse momentum $\hbar\mathbf{k}_\perp$, energy $\hbar\omega$
  and coordinate $z'$ will give rise to a field of the same energy and momentum at $z$. When then $z$ and $z'$ are only
   infinitesimally different, the only ways this can happen by classical reasoning is that the two are in fact exactly the same
   (corresponding to the $|z-z'|$-dependent source term) or that the field has been reflected off both walls once or more. This
   is what figure \ref{fig_refl} demonstrates.

%************* SECTION ******************
\section{Numerical investigation and temperature effects}\label{ch_numerics}

For our numerical calculations, we have used permittivity data for
aluminium, gold and copper supplied by Astrid Lambrecht (personal
communication). For ease of comparison, aluminium is used in
figures throughout; all variations aqcuired by replacing one metal
by another are of a quantitative, not qualitative nature, and are
not included here. In all our numerical investigations, we have
assumed non-magnetic media, i.e.\ $\mu_1=\mu_2=\mu_g=1$.

As an example of a dielectric, we have chosen teflon fluorinated ethylene propylene (teflon FEP) because its chemical and physical properties are remarkably invariant with respect to temperature. Permittivity data for teflon FEP are taken from \cite{chaudhury84}.

\begin{figure}
  \begin{center}
    \includegraphics[width=4in]{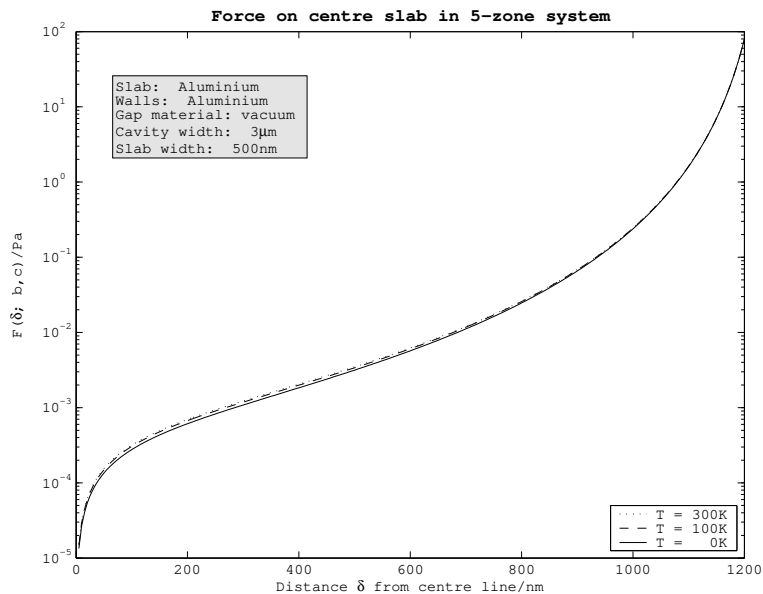}
    \caption{The force on an Al slab in a vacuum-filled cavity between Al walls. $\delta$ is the distance from the centre of the slab to the midline of the cavity. For negative $\delta$ one gets the antisymmetrical extension of the graph.}
    \label{fig_5zF}
  \end{center}
\end{figure}

Figure \ref{fig_5zF} shows the Casimir force acting on a relatively thick aluminium slab in a cavity as a function of $\delta$. For negative values of $\delta$ the situation is identical but the force has the opposite direction. We have chosen a gap width of 3$\mu$m and a slab thickness of 500nm. These values are not arbitrary: First, the relative temperature corrections of the Casimir force are predicted to be large at plate separations of 1-3$\mu$m, so a slab-to-wall distance in this region is desirable (here $h/2 = 1250$nm). Secondly, choosing the slab significantly thicker than the penetration depth of the EM field makes the five-zone geometry instantly comparable to the well-known three-zone Lifshitz geometry of two half-spaces; for slabs of a good metal thicker than $\sim 50$nm there is virtually no difference between the five-zone expression as derived above and that which one would acquire applying the standard Lifshitz expression to each gap in turn and
 finding the net force density on the slab as the difference between the two.

\begin{figure}
  \begin{center}
    \includegraphics[width=4in]{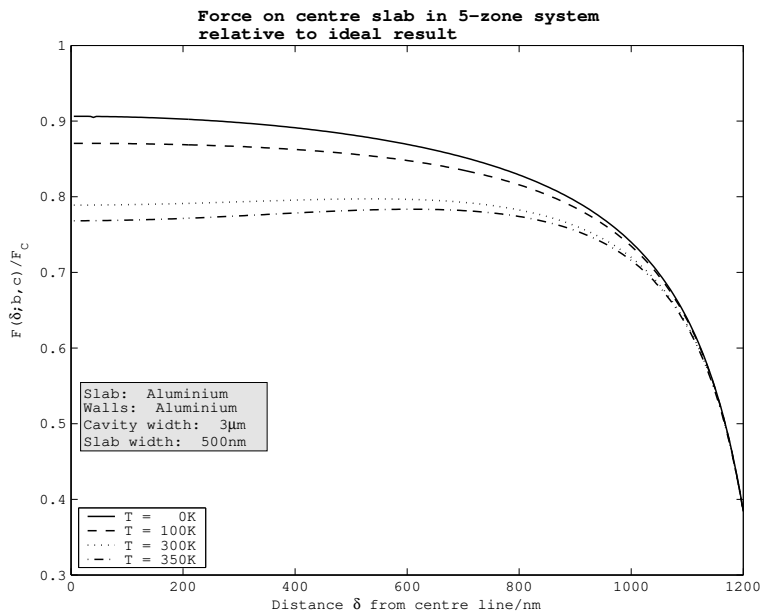}
    \caption{The force on an Al slab in a vacuum-filled cavity between Al walls relative to its value for ideally conducting slab and walls, equation (\ref{eq_Casimir_5z}). $\delta$ is the distance from the centre of the slab to the midline of the cavity. For negative $\delta$ one gets the symmetrical extension of the graph.}
    \label{fig_5zFbyFC}
  \end{center}
\end{figure}

Figure \ref{fig_5zFbyFC} shows the net vacuum pressure acting on the slab relative to Casimir's result for ideal conductors,
\be
  \label{eq_Casimir_5z}
  \mathcal{F}_\mathrm{C}= \frac{\hbar c\pi^2}{240}\left[\frac{1}{(h/2-\delta)^4}-\frac{1}{(h/2+\delta)^4}\right].
\ee In such a plot we see clearly how a slab and cavity set-up
might be suitable for measurements of temperature effects; whereas
such effects are small for very small separations, they grow most
considerable near the centre position where slab-to-wall distance
is in the order of a micrometre.

\begin{figure}
  \begin{center}
    \includegraphics[width=4in]{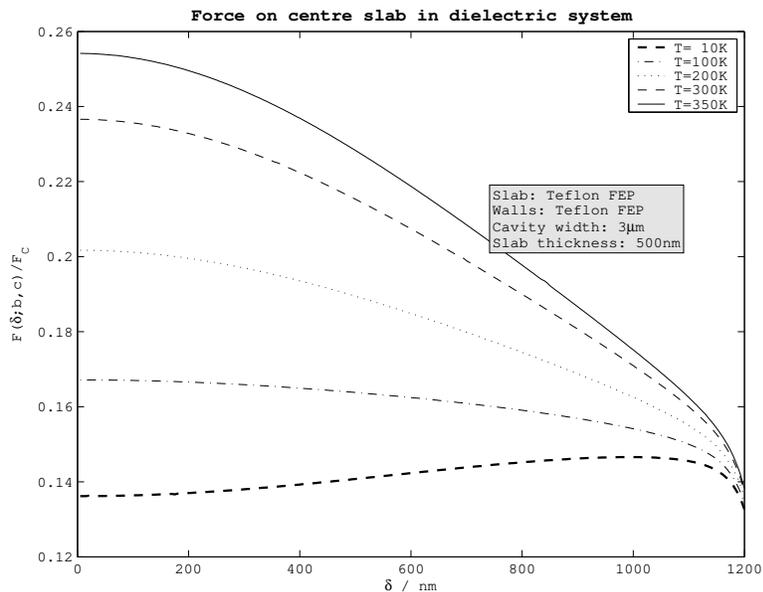}
    \caption{The force on a teflon FEP slab in a vacuum-filled cavity between teflon FEP walls relative to the result for ideally conducting slab and walls, equation (\ref{eq_Casimir_5z}). Note that dielectric properties are assumed constant with temperature.}
    \label{fig_5zTFEPFbyFC}
  \end{center}
\end{figure}

An altogether different result is obtained upon replacing metal with a dielectric in both walls and slab. In figure \ref{fig_5zTFEPFbyFC}
 the same calculation as in figure \ref{fig_5zFbyFC} has been performed with both slab and walls of teflon FEP. Casimir experiments
 using dielectrics were proposed by Torgerson and Lamoreaux \cite{torgerson04} where the use of diamond was suggested.

It is important to note here that we have not taken into account variations of the dielectric properties of teflon FEP with temperature;
 much as teflon FEP is renowned for its constancy in electrical and chemical properties over a large temperature range and is used in space
  technology for this very reason, one must assume there are corrections at extremely low temperatures. We shall not enter into a discussion on
  material properties here; the point to take on board is rather that temperature effects are found to be very large indeed near the centre
  position, a fact that does not change should the calculated values be several percent off. This strongly indicates that the use of dielectrics
  in Casimir experiments could be an excellent means of measuring the still controversial temperature dependence of the force.

We note furthermore that whilst for metals the force decreases with rising temperatures, the opposite is the case for the dielectric. Mathematically
 this is readily explained from e.g.\ (\ref{eq_5z_F(aT)}). Temperature enters into the expression in two ways; first, each term of the Matsubara sum
  has a prefactor $T$, secondly the distancing of the discrete imaginary frequencies increases linearly with $T$. The first dependence tends
   to increase the force with respect to $T$ whilst the other decreases it (bearing in mind that the integrand, which is proportional
    with $\exp(-\kg h)$, decreases rapidly with respect to $\zeta$ for $\zeta$ larger than roughly the $m=1$ Matsubara frequency). As temperature rises,
     thus, the higher order terms of the sum quickly become negligible, leaving the first few terms to
      dominate\footnote{The same phenomenon for increasing distances rather than temperatures is treated in \cite{brevik03}.}. In the high
      temperature limit, $m=0$ becomes the sole significant term and the force becomes proportional\setcounter{footnote}{0}\footnote{ For the three-layer Lifshitz set-up, the zero term and
       thus the force becomes proportional to $T/a^3$ where $a$ is the gap width, as shown formally in \cite{nesterenko04}.} to $T$. This is true for metals
      and dielectrics
   alike, but while the trend is seen at low temperatures for dielectrics, for metals the $T$-linear trend
typically becomes visible only at temperatures much higher than
room temperature. In metals the low (nonzero)
    frequency terms are boosted since $\epsilon_i\gg\epsilon_g$ for $\zeta$ much smaller than the plasma frequency,
     in which case reflection coefficients $|\Delta_{iq}|$ approximately equal unity. The first few Matsubara terms thus remain significant as temperature rises,
      countering the $T$-proportionality effect, at the same time as each $m>0$ term decreases in value as the Matsubara frequencies
       take higher values, allowing the resulting force to decrease with increasing temperature.

\begin{figure}
  \begin{center}
    \includegraphics[width=4in]{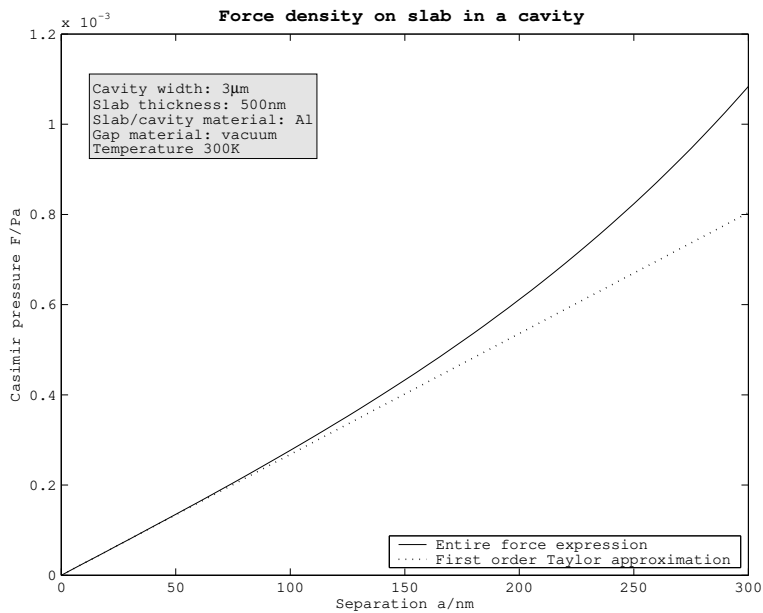}
    \caption{The Casimir force density on the slab of figure \ref{fig_5z} as a function of the distance $\delta$ from the centre of the slab
     to the cavity midline as compared to its first order Taylor expansion, equation (\ref{eq_Taylor}) at temperature 300K.}
    \label{fig_5z_Taylor}
  \end{center}
\end{figure}

Figure \ref{fig_5z_Taylor} shows the force acting on the slab in the previously described geometry (such as plotted in figure \ref{fig_5zF}) as well the first order Taylor expansion. The figure gives a rough idea as to the size of the central cavity region in which one may regard the force density as linear with respect to $\delta$. With the system parameters as chosen we see that, depending on precision one may allow oscillation amplitudes $\delta$ of several tens of nanometres, a length which is not small relative to the system.

\begin{figure}
  \begin{center}
    \includegraphics[width=4in]{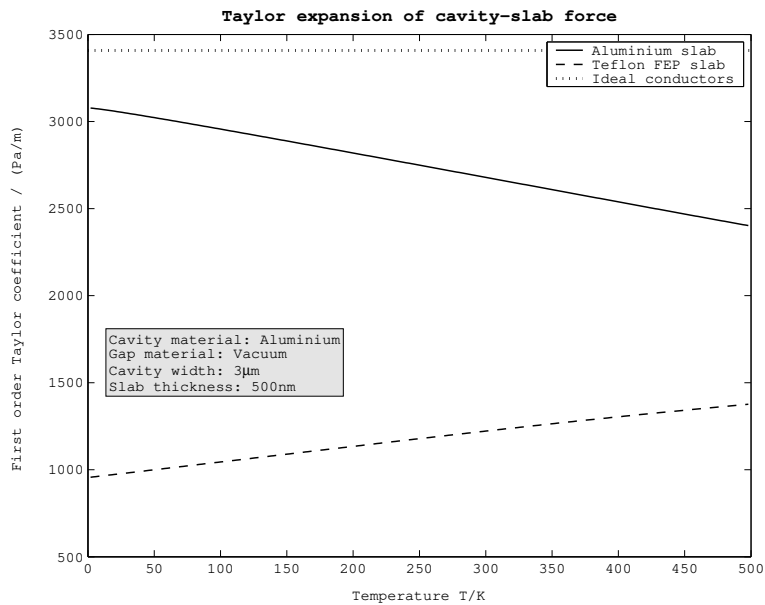}
    \caption{The first order Taylor coefficient of equation (\ref{eq_Taylor}) for aluminium and teflon FEP slabs in an Al cavity. The horizontal dotted line is the coefficient pertaining to the Casimir result for ideal conductors (both slab and walls), equation (\ref{eq_a1C}). One should note that the dielectric properties of the materials at extremely low temperatures are not known.}
    \label{fig_a1}
  \end{center}
\end{figure}

The first order Taylor coefficient itself has been calculated and plotted in figure \ref{fig_a1} for aluminium and teflon FEP slabs in an aluminium cavity. These are furthermore compared to Casimir's ideal result (\ref{eq_Casimir_5z}) whose first order Taylor coefficient is readily found to be
\be \label{eq_a1C}
  a_{1\mathrm{C}} = \frac{16\hbar c\pi^2}{15}h^{-5} \approx 3.3283\cdot 10^{-25}\mathrm{Nm}^2\cdot h^{-5}.
\ee

%************* SECTION ******************
\section{The effect of finite slab thickness}

As measurements of the Casimir force have become drastically more
accurate over the last few years, with researchers claiming to
reproduce theoretical results to within 1\%
\cite{lamoreaux05,decca05}, it is well worth asking whether any
\emph{theoretical} calculation may rightly claim such an accuracy.
A point of particular interest in this respect is the strong
dependence of the Casimir force on the permittivity of the media
involved. The permittivity data for aluminium, copper and gold
supplied by Lambrecht and Reynaud were calculated by using
experimental values for the susceptibility at a wide range of real
frequencies (approx.\ $1.5\cdot 10^{14}$rad/s$<\omega<1.5\cdot
10^{19}$rad/s), extrapolating towards zero frequency by means of
the Drude relation (for small $\omega<$ approx $1.5\cdot
10^{14}$rad/s). $\epsilon(\omega)$ was subsequently mapped onto
the imaginary frequency axis invoking Kramers-Kronig relations
numerically. Thus, although matching theoretical values (Drude
mode
 l) excellently for imaginary frequencies up to about $10^{15}$rad/s \cite{bentsen05}, the data have intrinsic uncertainties. Recently, Lambrecht and co-workers addressed the question of the uncertainty related to calculation of the Casimir force due to uncertainty in the Drude parameters used for extrapolation, found to add up to as much as 5\%, considerably more than the accuracy claimed for the best experiments to date \cite{pirozhenko06}. 

The effect of the ``leakage'' of vacuum radiation from one gap region to the other in our 5-zone geometry is worth a brief investigation in this context. Excepting the zero frequency term, it is unambiguous from e.g.\ the definition of $\kappa$ that when the slab is metallic, the factor $\exp(-2\kappa_2 b)$ is small compared to unity for sufficiently large values of $b$, due to the large values of $\epsilon_2$ for all important frequencies $\rmi\zeta$\footnote{The $m=0$ term is a subtle matter we shall not enter into here. For a recent review, see e.g.\ \cite{brevik06a} and references therein.}. Let us regard one of the gap regions, of index $\pm$. With some manipulation one may expand (\ref{d}) to first order in the factor $\exp(-2\kii b)$ to find the pertaining quantity
\begin{eqnarray}
  \frac{1}{d_q^\pm} &=& \frac{\Diq\Diiq \rme^{-2\kg a^{\pm}}}{1-\Diq\Diiq \rme^{-2\kg a^{\pm}}} -\nonumber \\
  &&\rme^{-2\kii b}\frac{\Diq(1-\Diiq^2)\rme^{-2\kg a^\pm}}{(1-\Diq\Diiq \rme^{-2\kg a^\pm})^2}\cdot\frac{\Diiq-\Diq\rme^{-2\kg a^\mp}}{1-\Diq\Diiq\rme^{-2\kg a^\mp}} + \mathcal{O}(\rme^{-4\kii b}).
\end{eqnarray}
The first term is immediately recognised as giving the Lifshitz expression for the Casimir attraction between two half-spaces of materials $1$ and $2$ separated by a gap of width $a^\pm$ and material $g$, and the second term is the first order correction due to penetration of radiation through the slab.

In terms of $\delta$ we may write in the case where $\exp(-2\kii b)\ll 1$ for all relevant frequencies (again subsequent to some manipulation) the force on the slab as $\mathcal{F}^T(\delta) \approx \mathcal{F}^T_\mathrm{L} + \Delta\mathcal{F}^T$ where
\[
  \mathcal{F}^T_\mathrm{L}(\delta;h)=\frac{k_BT}{\pi}\matSum \infint \dd k_\perp k_\perp \kg \sEM \frac{A_{q\mathrm{L}}\sinh 2\kg\delta}{B_{q\mathrm{L}}-A_{q\mathrm{L}}\cosh 2\kg\delta}
\]
is the result using the Lifshitz expression on both gaps and taking the difference; here
\be
  A_{q\mathrm{L}} \equiv 2\Diq\Diiq \rme^{-\kg h}\hspace{20pt}\mathrm{and}\hspace{20pt}  B_{q\mathrm{L}} \equiv 1+\Diq^2\Diiq^2\rme^{-2\kg h};
\ee
and
\begin{eqnarray}
  \Delta\mathcal{F}^T (\delta;h,b)&=& -\frac{k_BT}{\pi}\matSum \infint \dd k_\perp k_\perp \kg \sEM \rme^{-2\kii b}\times\nonumber\\
  &&\times\frac{A_{q\rmL}(B_{q\rmL}-\Diiq^2-\Diq^2\rme^{-2\kg h})\sinh 2\kg \delta}{(B_{q\mathrm{L}}-A_{q\mathrm{L}}\cosh 2\kg\delta)^2}.\label{eq_DeltaF}
\end{eqnarray}

\begin{figure}
  \begin{center}
    \includegraphics[width=4in]{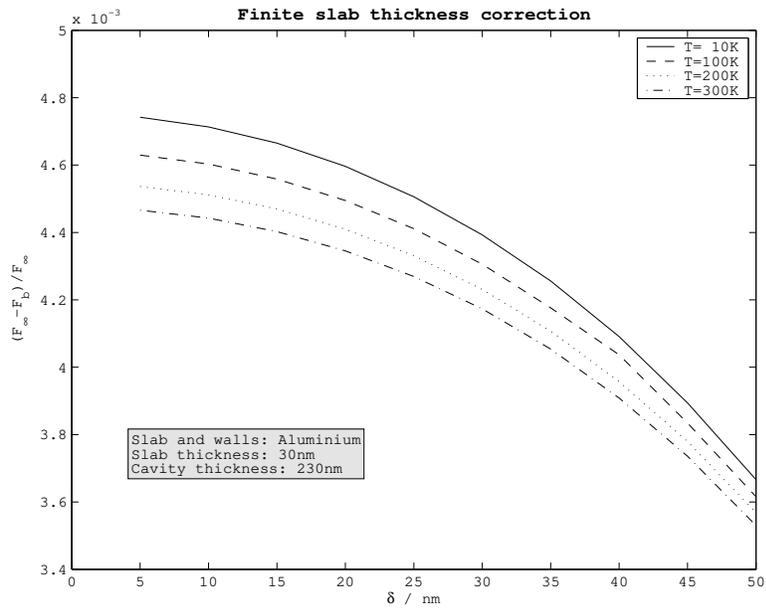}
    \caption{Thickness correction due to ``leakage'' of radiation through a thin slab in a small cavity. The absolute value of the correction is approximately exponentially decreasing with the thickness $b$.}
    \label{fig_thickness}
  \end{center}
\end{figure}

\begin{figure}
  \begin{center}
    \includegraphics[width=4in]{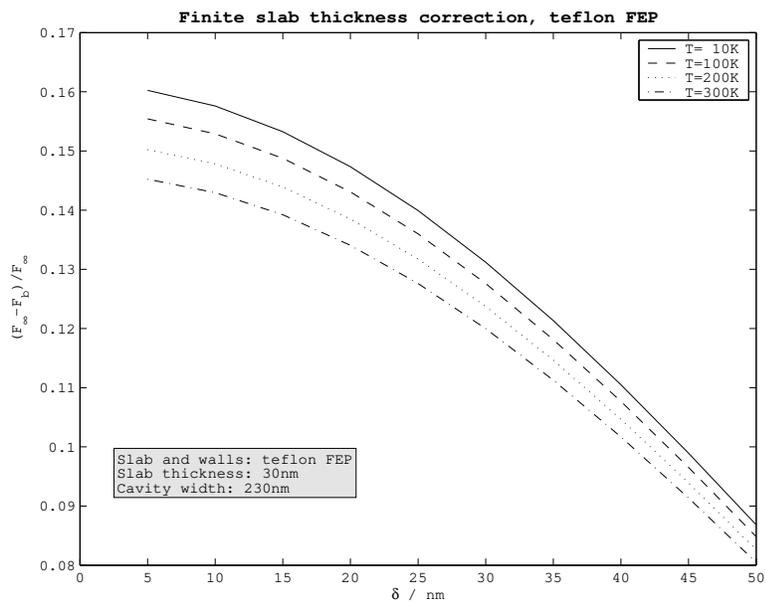}
    \caption{Thickness correction due for teflon FEP.}
    \label{fig_thicknessTFEP}
  \end{center}
\end{figure}

The factor $\exp(-2\kii b)$ and consequently the first order correction, is very sensitive with respect to even small
 changes in $\epsilon_2(\rmi \zeta)$. For very thin slabs ($b<50$nm) and small cavities, the correction could be in the order
  of magnitude of the currently claimed measurement accuracy. Furthermore, we see that the integrand of (\ref{eq_DeltaF}) depends
  on $\epsilon_2$ in an exponential way. In conclusion: to the extent that the thickness correction is of significance in an experimental
   measurement, exact knowledge of the permittivity as a function of imaginary frequency is of the essence. In such a scenario,
    approximate knowledge of the dispersion function could effectively limit our ability to even \emph{calculate} the force with the
    precision that recent experiments claim to reproduce theory \cite{decca05}.

In the case of dielectrics, the correction is almost two orders of magnitude larger and should be readily measurable. Experiments in a geometry involving dielectric plates of finite thickness might even be a possible means of evaluating the correctness of the dielectric function employed.

%%%%%%%%%%%%%%%%%%%%%%%%%%%%%%%% SECTION %%%%%%%%%%%%%%%%%%%%%%%%
\section{Conclusion and final remarks}

The main conclusion from the work presented is that from a theoretical point of view the five-zone setup (figure \ref{fig_5z}) as discussed could be ideal for detection of the temperature dependence of the Casimir force when the wall-to-slab distance is in the order of 1$\mu$m. One method as suggested is a measurement of the difference in the eigenfrequency of an oscillating slab in the absence and presence of a cavity.

When metal is replaced by a dielectric in slab and walls, relative temperature corrections become much larger, suggesting that using dielectrics whose dielectric properties vary little with respect to temperature be excellent for such measurements.

Our treatment of the effect of finite slab thickness shows that
the effect of finite thickness varies
 dramatically with respect to the properties of the materials involved, specifically $\epsilon$ and $\mu$.
  Much as the effect is generally quite small for metals, to the extent such effects do play a role even
  a moderately good estimate of their exact magnitude requires very accurate dielectricity data for the material in question.
  This is but one example of the more general point that the still considerable uncertainties associated with the best
   available permittivity data for real materials call for soberness in any assessment of our ability to numerically
   calculate Casimir forces with great precision.

   Finally, a couple of remarks: it ought to be pointed out  that our proposed
   method of investigating the thermal Casimir force via
   observing the oscillations of a slab in a cavity, can be
   classified as belonging to the subfield usually called the
   ``dynamic Casimir effect''. The use of mechanical microlevers has turned out to be very effective
   components for high sensitivity position measurements, of
   interest even in the context of gravitational waves detection.
   As for the basic principles of the method see, for instance
   Jaekel {\it et al.} \cite{jaekel02} with further references
   therein, especially Ref.~\cite{jaekel92}.  For more recent papers on microlevers, see
   Refs.~\cite{milonni04,karrai06,arcizet06,kleckner06}. 

   Moreover, we note the connection between our approach and the
   statistical mechanical approach of Buenzli and Martin
   \cite{buenzli05}. These authors computed the force between two
   quantum plasma slabs within the framework of non-relativistic
   quantum electrodynamics including quantum and thermal
   fluctuations of both matter and field. It was found that the
   difference in the predictions for the temperature dependence of
   the Casimir effect are satisfactorily explained by taking into
   account the fluctuations {\it inside} the material. Their
   predictions for the force are in agreement with ours.

\appendix
\newcommand{\sw}{\mathrm{sandwich}}
\newcommand{\op}{\mathrm{open}}
\newcommand{\Fsw}{\mathcal{F}_\sw}
\newcommand{\Fop}{\mathcal{F}_\op}

\section{Closed geometry versus open configuration}\label{appendix}

We will demonstrate briefly the two physical properties favouring a closed cavity configuration (figure \ref{fig_5z}) as compared to an open configuration in which a similarly oscillating plate is held in equilibrium by an external spring system. For the purpose of comparison we will disregard effects due to finite plate thickness, so that the net force experienced by a slab in a cavity is the difference between standard Lifshitz forces on both sides, whist that between plate and wall in a one-sided geometry (like figure \ref{fig_5z} but with the right hand wall removed) is simply the  Lifshitz force. For separations of some hundred nanometres or more, the Lifshitz force varies as $\FL(d)\propto d^{-4}$. Thus the net force on the slab in the sandwich geometry attached to a spring of spring contstant $k$ per unit transverse area is (we assume $k>\mathcal{F}(a)/\delta$ as before)
\begin{eqnarray}
  \Fsw &=& \FL(a+\delta) - \FL(a-\delta) - k\delta\nonumber\\
  &=& \frac{F(a)}{(1+\delta/a)^4} - \frac{F(a)}{(1-\delta/a)^4}-k\delta \nonumber\\
  &=& -\left[ka-8\left|\FL(a)\right|\right]\frac{\delta}{a}+40\left|\FL(a)\right|\frac{\delta^3}{a^3}+...
\end{eqnarray}
where $a=h/2$ is here the distance from slab to wall in equilibrium position ($\FL(d)<0$).

\begin{figure}
  \begin{center}
    \includegraphics[width=3.5in]{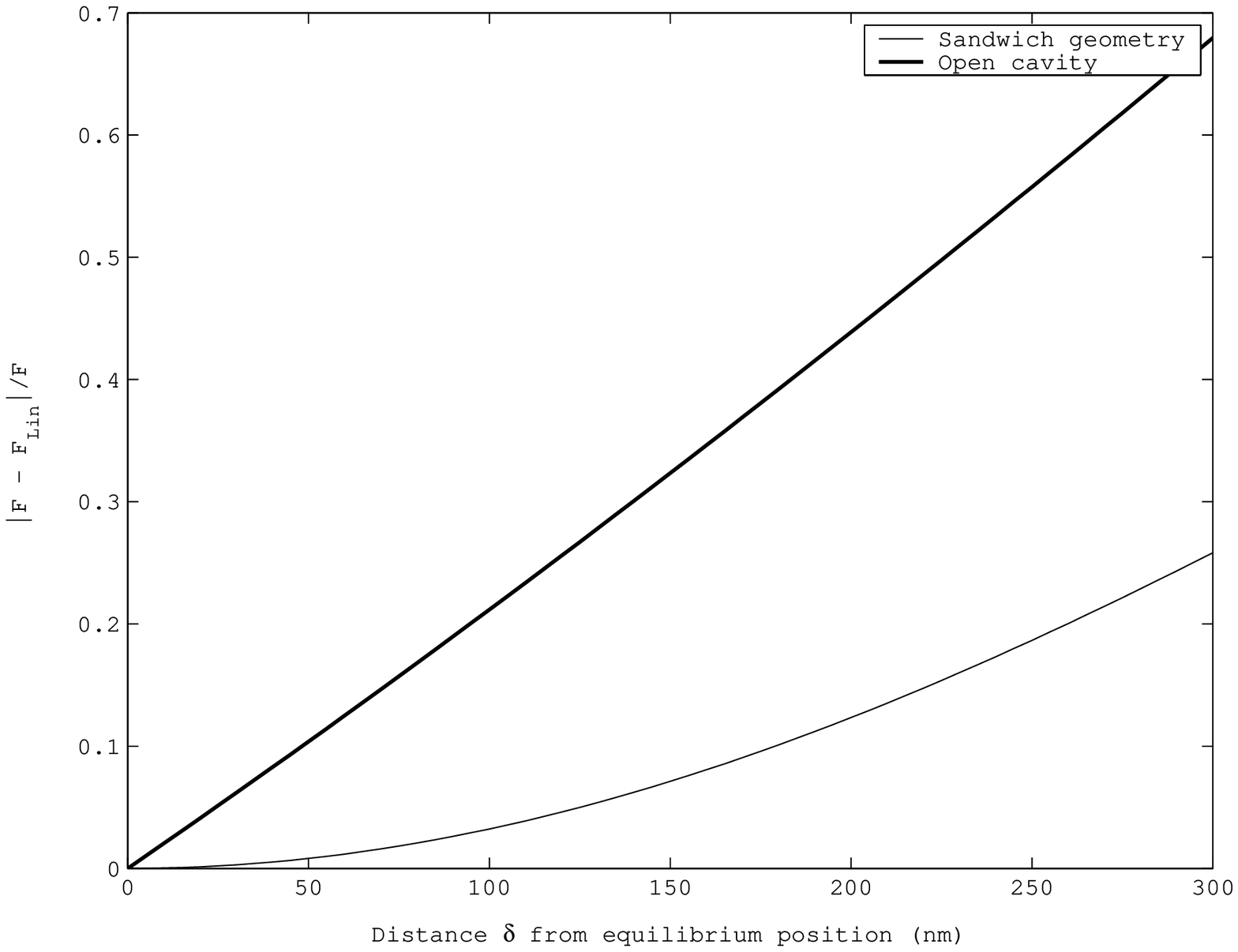}
    \caption{The relative correction to the linear Taylor expansion, $\mathcal{F}_\mathrm{Lin}$, of the Casimir force near equilibrium position for the open and closed geometries plotted for positive $\delta$. Calculations assume aluminium plate and walls, equilibrium plate-to-wall separation of $a=$1250nm in both configurations and temperature 300K. For the sandwich geometry, $\mathcal{F}$ is given by (\ref{eq_5z_F(aT)}) whilst in the open geometry $\mathcal{F}=\FL(a+\delta)-\FL(a)$ with $\FL$ the standard Lifshitz expression for the attraction between two half-spaces.}
    \label{fig_5z4z}
  \end{center}
\end{figure}

Now consider an open configuration in which a plate is held in equilibrium by an external spring, also of spring constant $k$ per unit transverse area. Assume that the forces are in equilibrium when the plate is a distance $a$ from the cavity wall. The net force on the plate is
\begin{eqnarray}
  \Fop &=& \FL(a+\delta) - \FL(a) - k\delta \nonumber\\
  &=& -[ka - 4|\FL(a)|]\frac{\delta}{a} - 10 |\FL(a)|\frac{\delta^2}{a^2} +...
\end{eqnarray}

There are thus two properties that favour the closed geometry. First, the first-order perturbation of the spring constant is twice as large and second, that the leading order correction to the harmonical approximation ($\mathcal{F}^T\propto \delta/a$) is cubical whist it is quadratic for the open configuration\footnote{Note that the specific assumption $\FL(d)\propto d^{-4}$ is not necessary for either of these results; they pertain almost exclusively to geometry. A more general power $\FL(d)\propto d^{-\sigma}$, $\sigma>0$, say, gives the same properties.}. The closed geometry thus allows considerably larger deviations from equilibrium position at a given accuracy without taking non-harmonic effects into account. In figure \ref{fig_5z4z} this is demonstrated by plotting the relative nonharmonic correction as a function of $\delta$ at a separation 1250nm. In accordance with our results, the relative correction $(\mathcal{F}-\mathcal{F}_\mathrm{Lin})/\mathcal{F}$ is approximately linear for an open geometry ($\approx -\frac{5}{2}\frac{\delta}{a}$) and approximately quadratic for a sandwich ($\approx 5\frac{\delta^2}{a^2}$).

\section*{Acknowledgements}

Permittivity data for aluminium, gold and copper used in our calculations were kindly made available to us by Astrid Lambrecht and Serge Reynaud.
 We furthermore thank Valery Marachevsky for stimulating discussions and
 input, and Mauro Antezza for valuable correspondence.

\end{document}